\RequirePackage[2020-02-02]{latexrelease}
\documentclass[aps,prr,superscriptaddress,manuscript,longbibliography]{revtex4-1}

\usepackage{epsfig}
\usepackage{times}
\usepackage{xcolor}

\begin{document}
		
\title{When faster rotation is harmful: the competition of alliances with inner blocking mechanism}

\author{Attila Szolnoki}
\email{szolnoki.attila@ek-cer.hu}
\affiliation{Institute of Technical Physics and Materials Science, Centre for Energy Research, P.O. Box 49, H-1525 Budapest, Hungary}

\author{Xiaojie Chen}
\email{xiaojiechen@uestc.edu.cn}
\affiliation{School of Mathematical Sciences, University of Electronic Science and Technology of China, Chengdu 611731, China}

\begin{abstract}
Competitors in an intransitive loop of dominance can form a defensive alliance against an external species. The vitality of this super-structure, however, is jeopardized if we modify the original rock-scissors-paper-like rule and allow that the vicinity of a predator blocks stochastically the invasion success of its neighboring prey towards a third actor. To explore the potential consequences of this multi-point interaction we introduce a minimal model where two three-member alliances are fighting but one of them suffers from this inner blocking mechanism. We demonstrate that this weakness can be compensated by a faster inner rotation which is in agreement with previous findings. This broadly valid principle, however, is not always true here because the increase of rotation speed could be harmful and results in series of reentrant phase transitions on the parameter plane. This unexpected behavior can be explained by the relation of the blocked triplet and a neutral pair formed by a triplet member with an external species. Our results provide novel aspects to the fundamental laws which determine the evolutionary process in multi-strategy ecological systems.
\end{abstract}

\maketitle

\section{Introduction}
\label{intro}
A stronger actor in an ecological system cannot be safe because a third type of competitor may dominate it. Ironically, the originally ``weaker'' player could also be the predator of the last one, which establishes a cyclic dominance among these three rival participants. Such intransitive relation, that is captured in a rock-scissors-paper game~\cite{szolnoki_jrsif14,dobramysl_jpa18}, can be detected in several living systems, starting from bacteria~\cite{shibasaki_prsb18,garde_rsob20,liao_ncom20} and plants~\cite{cameron_jecol09,lankau_s07} to animals~\cite{guill_jtb11,sinervo_n96}, or even among humans who follow different strategies in a social dilemma situation~\cite{hauert_s02,lee_hw_csf24,canova_jsp18,szolnoki_srep21,li_wj_amc22,szolnoki_pre15}. Over the last decade several excellent works have studied the behavior of such systems where cyclical dominance has a central role on the emerging pattern formation~\cite{avelino_pre19b,mir_pre22,tainaka_ei21,yoshida_srep22,yang_csf23,baker_jtb20,mobilia_g16,avelino_epl23,park_c19c,serrao_epjb21}.

If we step onto a higher level, such triplet~\cite{szabo_jpa05,szabo_pre07}, or even a larger loop~\cite{mir_pre22,park_csf23,brown_pre19}, may also be considered as an entity because their members can form a defensive alliance against an external species or alternative group~\cite{mao_yj_epl18}. It is an interesting question whether we can identify general dynamical rules which determine the vitality of such formation, especially in the case when equally strong groups compete for space in an evolutionary process. For example, a group formed by less members is generally more effective than a larger loop~\cite{szolnoki_amc23,de-oliveira_csf22,szolnoki_srep23}. Or, a loop having faster inner rotation is more powerful than the alternative formation where the average invasion is less intense~\cite{perc_pre07b,szolnoki_epl15}. The latter, however, is not necessarily a decisive factor because a members of a loop are not always equally strong~\cite{tainaka_pla95,avelino_epl21,juul_pre13}. As a consequence, a group which is formed by equally strong partners can surpass a rival group of diverse members where the average inner invasion is faster~\cite{szolnoki_epl20}. 

The efficiency of a defensive alliance may be affected by alternative mechanisms, too. In particular, the actual invasion strength between a predator-prey pair could also be influenced by the close vicinity of a third party~\cite{szolnoki_njp15}. In a microbial community, as a specific example, an antibiotic-degrading species attenuates the inhibitory interactions between two other species~\cite{kelsic_n15}. The consequence of such multi-point interaction~\cite{palombi_epjb20} could be specially interesting in a cyclically dominated loop because it does not simply modify the inner dynamics of the loop-members, but it can also change the invasion strength toward external species. 

To explore how such kind of extension varies the vitality of a defensive alliance, we introduce a minimal model where two three-member loops compete. One of them still represents the traditional rock-scissors-paper-like interaction where during a potential invasion only the state of involved pair counts. In the alternative triplet, however, we introduce a blocking mechanism. More precisely, staying at the rock-scissors-paper example, a rock will only beat scissors with a limited probability if there is a paper in the nearest neighborhood of the actor representing rock state.  Similar inhibition is also applied when a rock tries to beat an external species, but the neighboring paper blocks it. Summing up, the blocking mechanism is valid for not just the inner dynamics, but also hurts the external invasion capacity of the affected triplet. 

At first glance, the additional blocking rule has only negative consequence on the vitality of the mentioned triplet, but this seemingly reasonable conclusion is not always valid. As we show in this work, a faster inner invasion is capable to compensate the shortage of the alliance. Interestingly enough, a faster rotation among triplet members could also be harmful, which has never been observed. This unexpected system behavior can be explained by the relation of the blocked triplet and a pair formed by an inner member and an external species. As a consequence, the concept of defecting alliance is more subtle then we thought earlier.

The rest of this paper is organized as follows. In Sec.~\ref{def} we define the details of the minimal six-species model which is capable to test the potential consequence of the blocking mechanism. Here we also present the first observation obtained in a limited three-species subsystem which establishes the preliminary expectations about the modified dynamical rule. Our main results are summarized in Sec.~\ref{results} where we systematically visit the full parameter space of the minimal model. This involves to survey series of phase diagrams obtained at some representative combination of model parameters. We here also give arguments which explain the unexpected system behavior at specific parameter values. Finally, we conclude with the summary of the results and a discussion of their implications in Sec.~\ref{conclusion}.

\section{Blocking the invasion by a neighbor}
\label{def}

In our minimal ecosystem species, marked by {\it 0}, $\dots$, {\it 5}, are distributed on a $L \times L$ square lattice where periodic boundary conditions are used. Each point is occupied by one of the species, no empty site is allowed. During an elementary step we select two neighboring sites randomly and if they are occupied by a predator-prey pair then the predator invades the position of the prey with a certain probability. To execute a full Monte Carlo (MC) step we repeat this procedure $N = L^2$ times. 

\begin{figure}
	\centerline{\epsfig{file=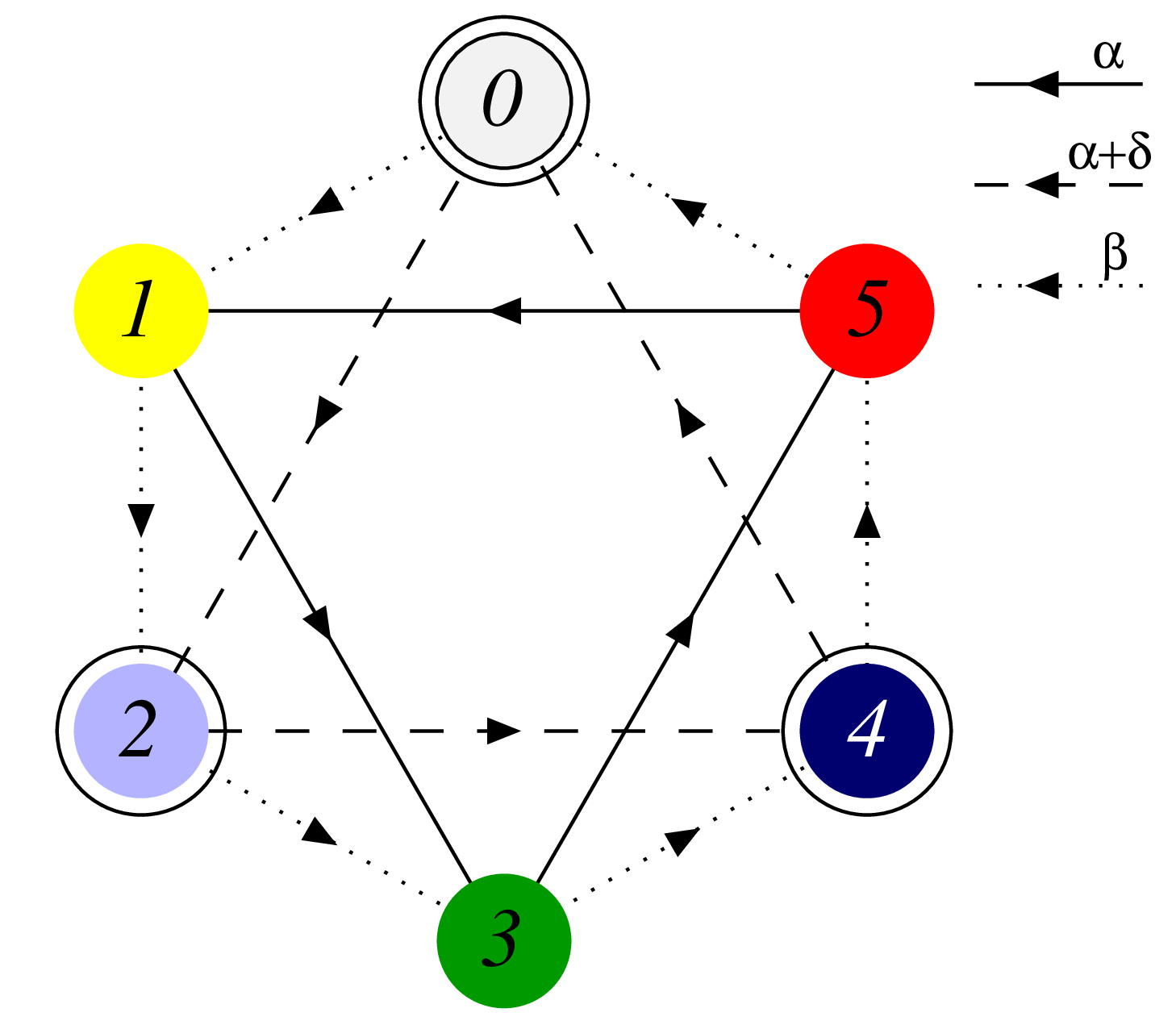,width=7.0cm}}
	\caption{\label{model}Food-web characterizing the microscopic dynamics in our six-member ecological system. Arrows indicate the direction of invasion between species. Two triplets form defending alliances, whose members are in a rock-scissors-paper-like relation. While the inner invasion probability in the {\it 1}$\to${\it 3}$\to${\it 5}$\to${\it 1} loop is $\alpha$, the same probability in the {\it 0}$\to${\it 2}$\to${\it 4}$ \to${\it 0} circle is $\alpha+\delta$. Importantly, the members of the latter triplet can block each other with probability $\gamma$. If, for instance, there is a species {\it 4} in the neighborhood of species {\it 0} then the {\it 0}$\to${\it 2} or the {\it 0}$\to${\it 1} invasion is blocked with probability $\gamma$. The invasion probability between the members of competing triplets is $\beta$.}
\end{figure}

The possible predator-prey interactions are summarized in Fig.~\ref{model}. As the food-web illustrates, practically we have two three-member loops whose members are in a rock-scissors-paper relation with each other. For instance, the odd-labeled species beat each other cyclically with probability $\alpha$. The inner invasion rate in the alternative loop, formed by even-labeled species, is higher and characterized by probability $\alpha+\delta$. Here parameter $\delta$ describes the excess invasion level which provides a faster inner invasion to these species. On the other hand, there is a three-point interaction in the latter loop because the presence of a third party can block the invasion between the original predator-prey pair. As an example, in a normal case species {\it 0} invades a neighboring species {\it 2} with probability $\alpha+\delta$. But this invasion is blocked with probability $\gamma$ if there is a species {\it 4} in the neighborhood of species {\it 0}. Importantly, this blocking mechanism is valid not only for the inner invasions of the triplet, but also for the process between the competing triplets. Staying at the previously mentioned example, the invasion between species {\it 0} and {\it 1} is also blocked with probability $\gamma$ in the vicinity of species {\it 4}. Last, the invasion strength between the rival groups is equally strong which is characterized by parameter $\beta$.

\begin{figure*}
	\centerline{\epsfig{file=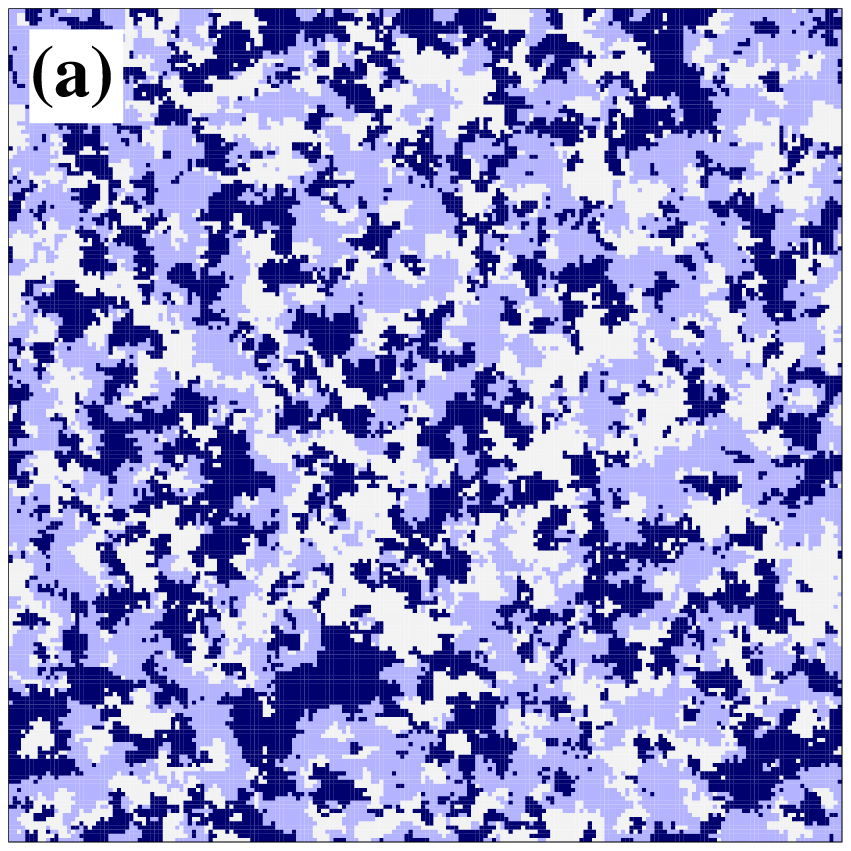,width=3.0cm}\epsfig{file=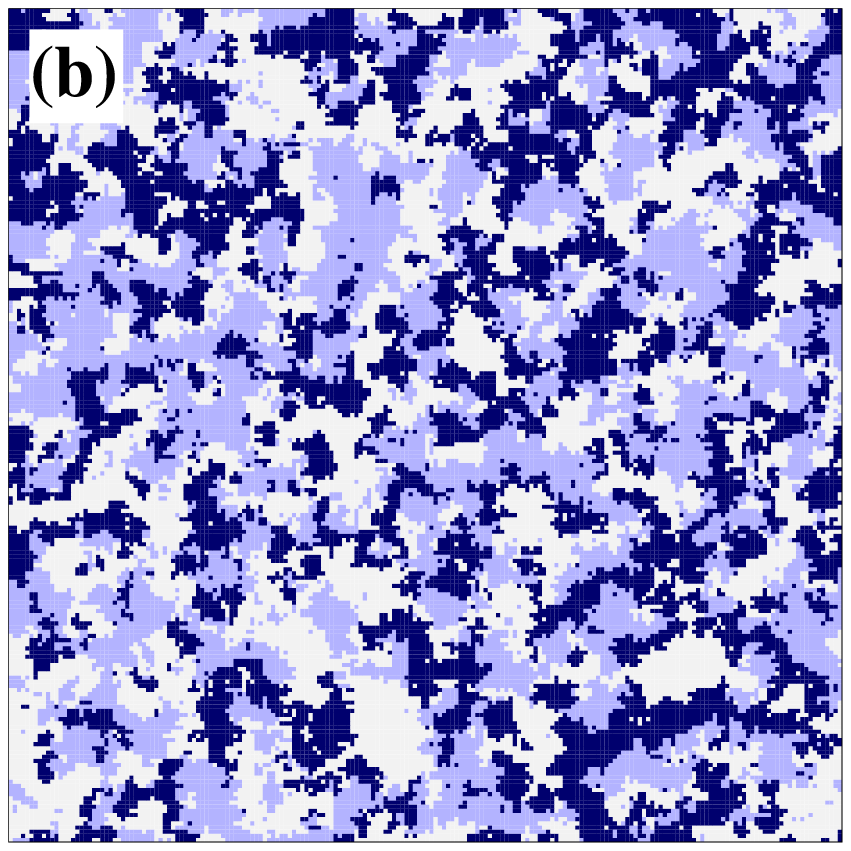,width=3.0cm}\epsfig{file=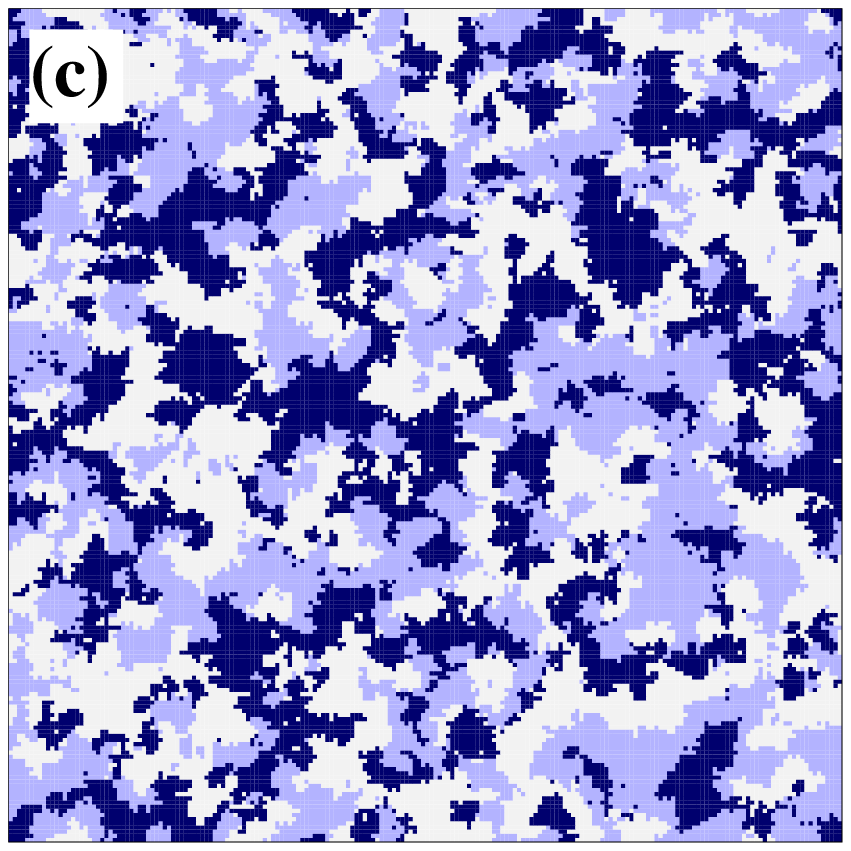,width=3.0cm}\epsfig{file=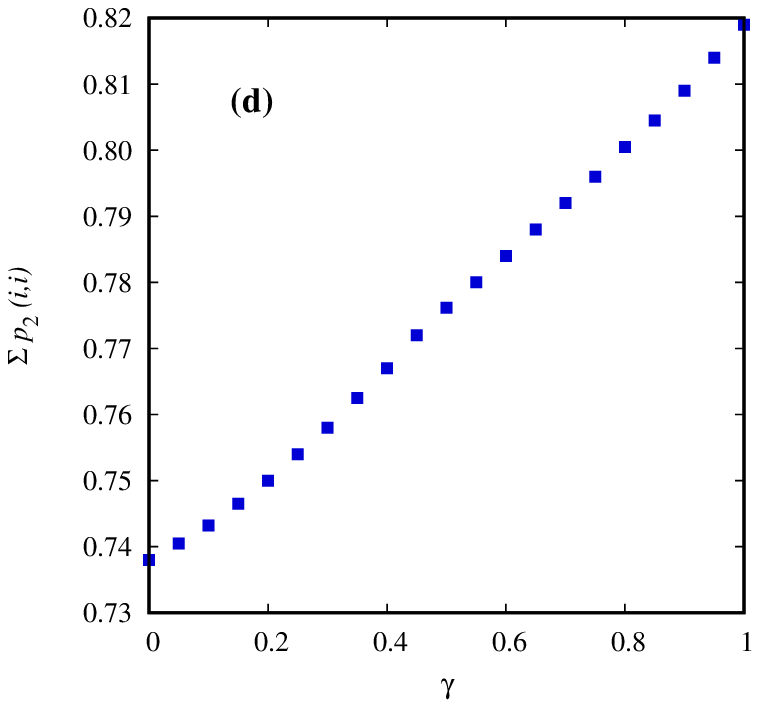,width=4.5cm}}
	\caption{\label{rsp}Stationary pattern of a three-species rock-scissors-paper system with blocking mechanism. In this sub-system only even-labeled species are present where the color code is identical to those we used in Fig.~1. Panels~(a) to (c) show spatially distribution of competing strategies obtained at $\gamma=0$, 0.5, and 1, respectively. The blocking mechanism effectively increases the typical domain size of competing species. Panel~(d) shows the two-point probability of identical species in dependence of $\gamma$ blocking rate.}
\end{figure*}

The above described blocking mechanism may explain why we use faster invasion in the even-labeled triplet because the external blocking is disadvantageous to them. There is, however, another aspect which also justifies why we need a faster invasion for a chance of surviving in the circle where blocking in the inner invasions may happen. To understand it, we present some representative patterns obtained at three different $\gamma$ values in a subsystem where only the even-labeled species are present. These are shown in the first three panels of Fig.~\ref{rsp} where the $\gamma$ values are 0, 0.5, and 1 respectively. Evidently, the symmetry between the species are not broken, but these plots suggest that the average domain size increases for larger $\gamma$ values. This effect can be confirmed quantitatively if we measure the stationary value of two-point probability of identical species which is proportional to the characteristic length of domain sizes. More precisely, we measure $\sum_i p_2(i,i)$ for $i=0,2,4$ in the stationary state obtained at a specific value of $\gamma$. Panel~(d) in Fig.~\ref{rsp} shows that this value, hence the average domain size, grows practically linearly with $\gamma$. 

Based on the above described observations we can conclude that stronger inner blocking mechanism produces larger domains among the species who are involved in the modified dynamics. According to our previous experience obtained for the competition of unequal triplets, larger mono-domains could be the Achilles-point of a defending alliance because a large homogeneous domain is always an easy target for a hostile alliance based on cyclic rotation~\cite{szolnoki_epl20}. Therefore, we expect that the group of even-labeled species, who may block each other via multi-point interactions, is generally weaker than the alternative blocking-free alliance. In this way we need to ``support'' the former group by allowing faster inner invasion because speedy exchange among group members generally helps an alliance to survive~\cite{perc_pre07b}.

To explore the phase diagrams for all model parameters we need to use linear system size ranging from $L=400-6000$ because finite-size problems can easily be detected at the vicinity of phase transition points. For instance, when the stationary fraction of a certain species is low then the extinction of these species may easily happen at small system sizes. Therefore only those states should be considered valid at specific parameter values where the portions of species become independent of system size as $L$ is increased. Beside random starting configurations we also use prepared initial states where different patches of sub-solutions are separated in space at the beginning and after their proper fight is monitored~\cite{szolnoki_njp16,szolnoki_njp18b}. In this way we can reach the stationary solution that is valid in the large-size limit more reliably. The relaxation time depends on the system size, the initial configuration, and also on the distance from transition point. The typical relaxation time is between $2\cdot10^4-5\cdot10^6$ MC steps.

\section{Results}
\label{results}

Our model has four parameters, $\alpha, \beta, \gamma,$ and $\delta$, therefore we need a systematic study to explore the typical system behavior in the sea of parameter space. Since probability $\beta$ characterizes the interaction strength between competing alliances, we can distinguish three major cases from this aspect. Namely, we study the case when the interaction between the competing triplets is strong, intermediate, or weak. The other criterion which separates system behavior onto different subsections is the inner invasion strength characterizing the group of odd species. Here we also distinguish three classes when the $\alpha$ value is low, intermediate or high. The typical and representative system behavior are presented on the plane of the remaining two parameters, $\delta$ and $\gamma$. As noted, $\gamma$ characterizes the intensity of blocking mechanism in the alliance formed by even-labeled species, while $\delta$ denotes the excess rate in their inner invasion. 

\begin{figure*}
	\centerline{\epsfig{file=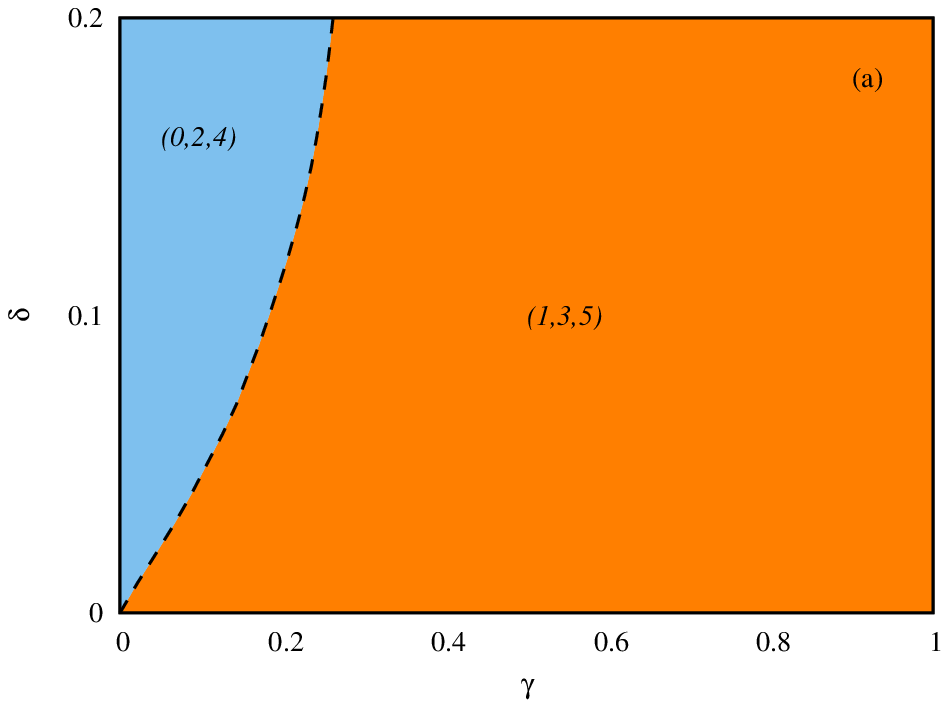,width=5.1cm}\epsfig{file=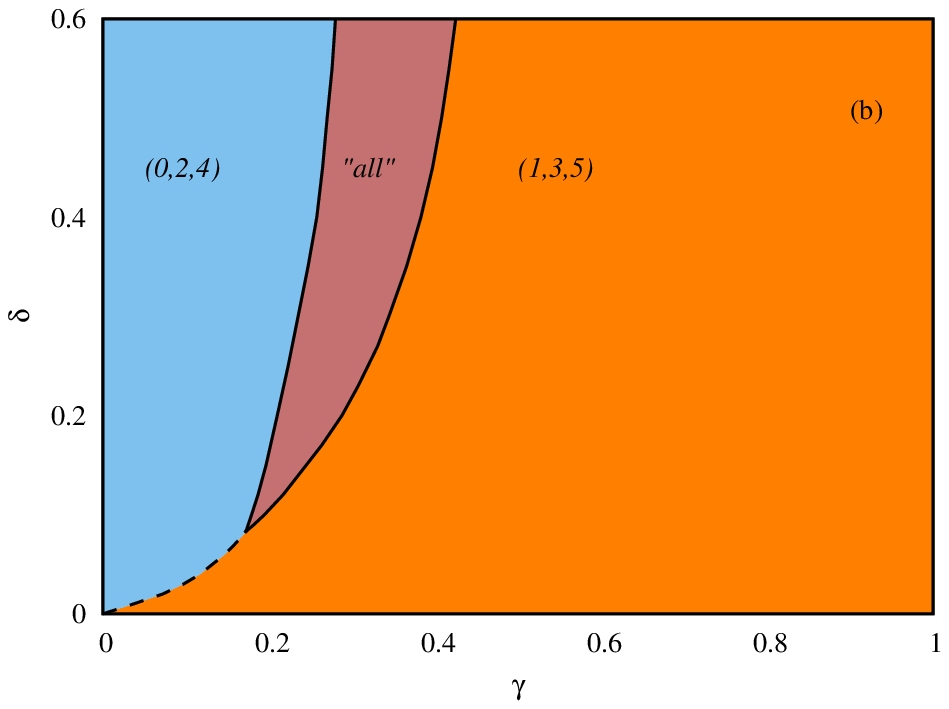,width=5.1cm}\epsfig{file=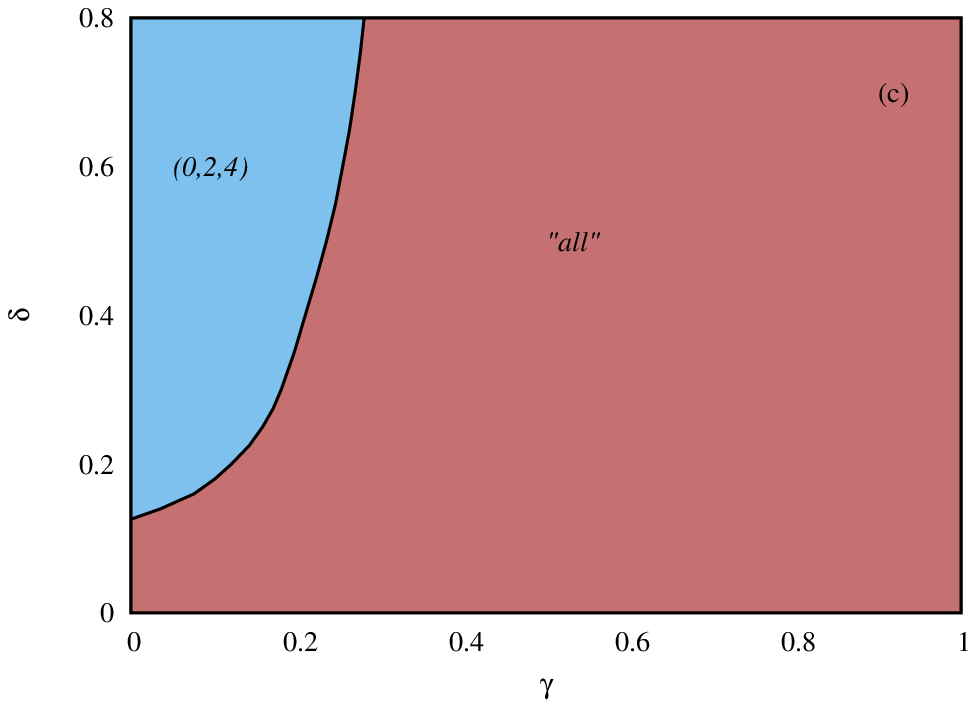,width=5.1cm}}
	\caption{\label{strong}Phase diagrams for strong interaction between rival triplets ($\beta=1$). Horizontal axis shows the $\gamma$ blocking strength while vertical axis denotes the $\delta$ extra inner invasion rate in the even-labeled group. Panels~(a) to (c) respectively show cases when the inner invasion strength in the odd group is strong ($\alpha=0.8$), medium ($\alpha=0.4$), or weak ($\alpha=0.2$). Light blue (orange) marks the parameter area where even (odd) species prevail, while pink shows the area where all six species survive. Dashed (solid) lines denote the position of discontinuous (continuous) phase transition points.}
\end{figure*}

Our first results, shown in Fig.~\ref{strong}, were obtained when the interaction between the triplets are strong. The three panels summarize the system behavior for $\alpha=0.8, 0.4$, and 0.2, where the inner invasion in odd-labeled loop is strong, intermediate or weak. Evidently, the probability $\alpha+\delta$ cannot exceed 1, which explains the different maximal $\delta$ values for each panels.

Panel~(a) shows the case when interactions in general are strong. As expected, when blocking mechanism is reasonable then the weakened even-labeled triplet cannot compete with the rival alliance and the group of ({\it 1+3+5}) species dominate the large-$\gamma$ region. Interestingly enough, however, the handicap of even-labeled group can be compensated by increasing their inner invasion. As we increase $\gamma$ blocking strength, we need larger $\delta$, faster rotation among ({\it 0+2+4}) species, to beat the stronger group. There is a discontinuous phase transition as we increase $\gamma$ which separates the domains where either even-labeled or odd-labeled triplet dominates the whole system. The character of this phase transition will be discussed in detail later.

At intermediate $\alpha$, shown in panel~(b), the border between the ({\it 0+2+4}) and ({\it 1+3+5}) phases become less sharp and there is a $\gamma$ interval when all six species survive. We marked this phase as ``{\it all}'' in the phase diagram. Finally, when the inner rotation is low in the odd-labeled loop, the orange domain of this state completely disappears, no mater the invasion of their members is never blocked. This cannot be said about the ({\it 0+2+4}) group, still, they can dominate in the low $\gamma$ - high $\delta$ corner in panel~(c) because fast inner rotation recovers their individual weakness. For all other parameter values all six species coexist represented by the pink ``{\it all}'' domain. The latter can be understood, because the low general invasion due to small $\alpha$ makes the three-member loops hazardous. The large $\beta$ value, on the other hand, makes the {\it 0}$\to${\it 1}$\to\dots\to${\it 5}$\to${\it 0} loop viable. In this way the latter formation prevails no matter it requires the simultaneous presence of more members.

Next we present two panels to illustrate the diverse character of phase transitions observed in the diagrams of Fig.~\ref{strong}. The sum of the fractions of even-labeled species is a good order parameter to characterize the actual state of the system. When it is 1 then only ({\it 0+2+4}) species survives. When it is zero then alliance formed by ({\it 1+3+5}) species prevails. In between these extreme values the system is in a state where all six species coexist.

\begin{figure}
	\centerline{\epsfig{file=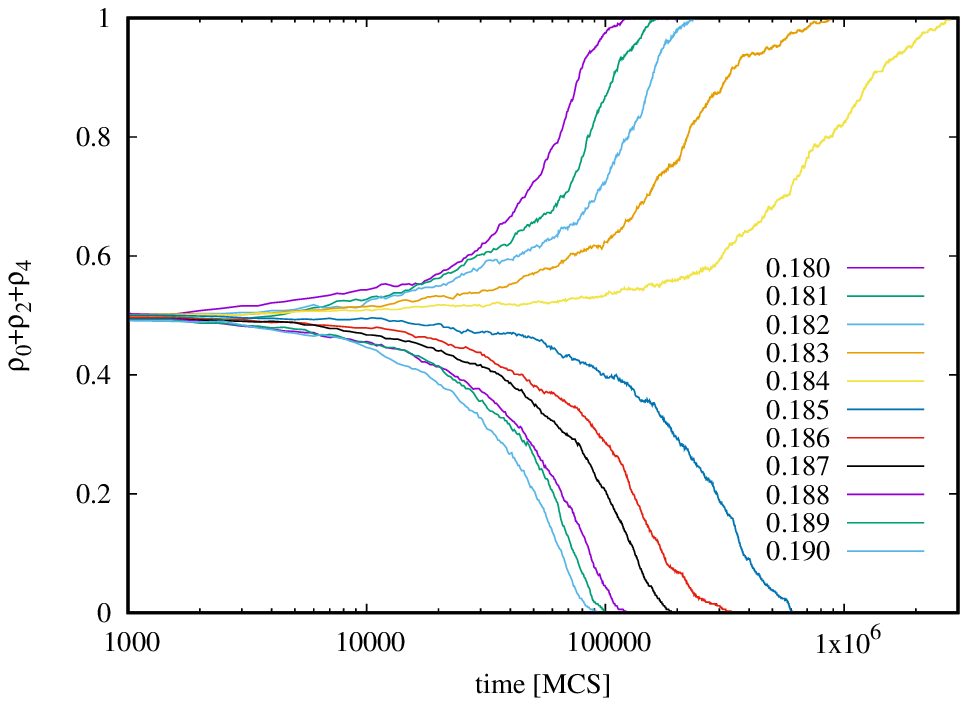,width=8.0cm}\epsfig{file=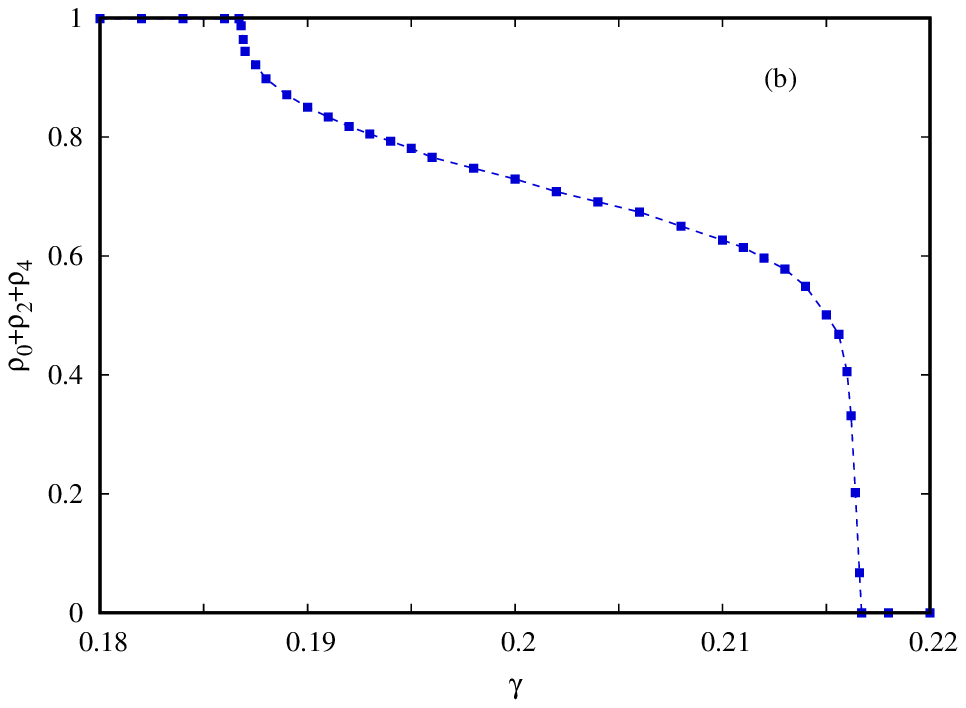,width=8.0cm}}
	\caption{\label{1_2}Panel~(a): Time dependence of the order parameter obtained for different $\gamma$ values as indicated in the legend at fixed $\beta=1$, $\alpha=0.8$, $\delta=0.1$ parameter values. The system evolves into either ({\it 0+2+4}) or ({\it 1+3+5}) state, and the change is sudden between $\gamma=0.184$ and 0.185. The linear system size is $L=800$ and we averaged over 100 independent runs. Panel~(b): Stationary state of the order parameter in dependence of $\gamma$ obtained at fixed $\beta=1$, $\alpha=0.4$, $\delta=0.12$. These values are independent of system size in the large-size limit.} It suggests that there is an intermediate phase where all six species coexist. Line is just to guide the eye.
\end{figure}

Figure~\ref{1_2}(a) illustrates a situation when we launch the evolution from a state where initially half of the space is occupied by ({\it 0+2+4}) species while the other half is filled by ({\it 1+3+5}) species. In this way we can monitor the fight of the alliances directly. As the panel shows, there is a sharp difference between the destinations as we gradually increase the $\gamma$ value while all other parameters are fixed. For smaller $\gamma$, when the blocking mechanism is moderate, the ({\it 0+2+4}) triplet can enjoy the advantage of faster inner invasion, while above a threshold $\gamma_c=0.1845(3)$ the negative consequence of blocking cannot be compensated anymore. This explains why we have discontinuous transition between ({\it 0+2+4}) and ({\it 1+3+5}) phases. As a technical note, this phenomenon can be observed at any system size, the only difference is the requested relaxation time to reach the final destination grows by increasing $L$.

Figure~\ref{1_2}(b) shows a completely different situation. Here the order parameter remains between zero and one in the stationary state stably signaling the emergence of a new phase where all six species are present. More precisely, if the system size is large enough then the value of order parameter becomes size-independent. Evidently, even smaller system size, like $L=400$, is enough to avoid finite-size effects if $\gamma$ is far from the critical transition point, while we need larger system sizes in the vicinity of $\gamma_c$. The results shown in Figure~\ref{1_2}(b) are valid in the large-size limit. As a conclusion, the transitions between the ``{\it all}'' and three-member phases are continuous.

\begin{figure*}
	\centerline{\epsfig{file=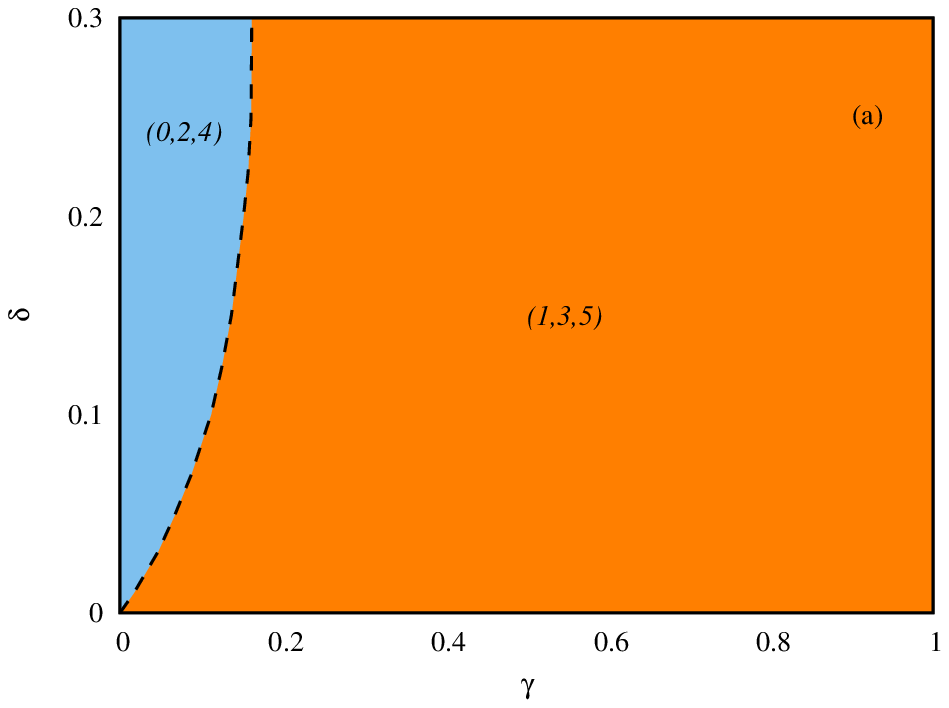,width=5.1cm}\epsfig{file=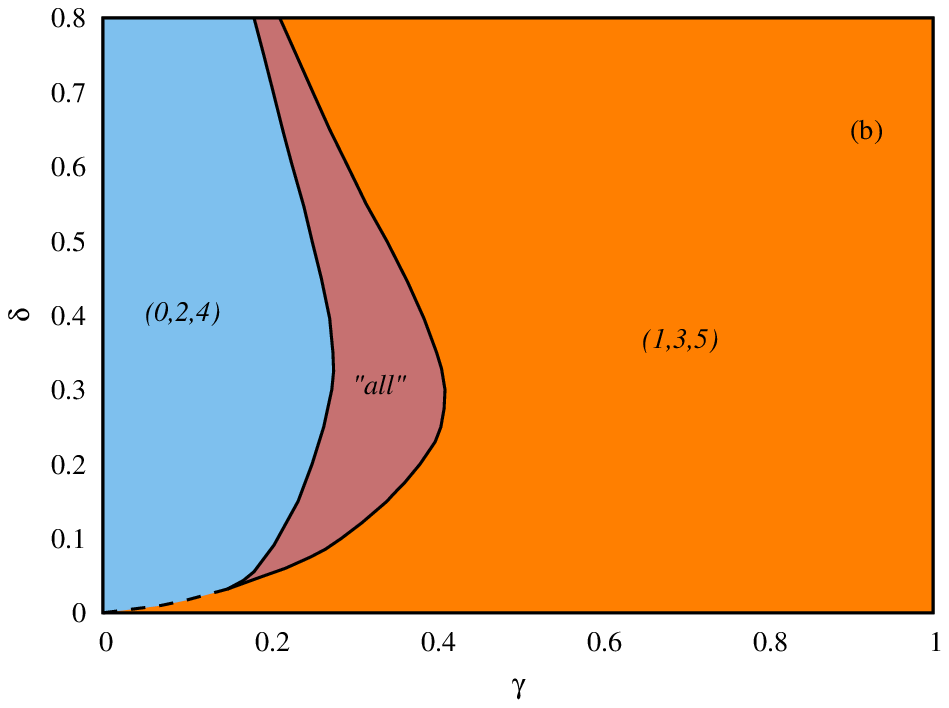,width=5.1cm}\epsfig{file=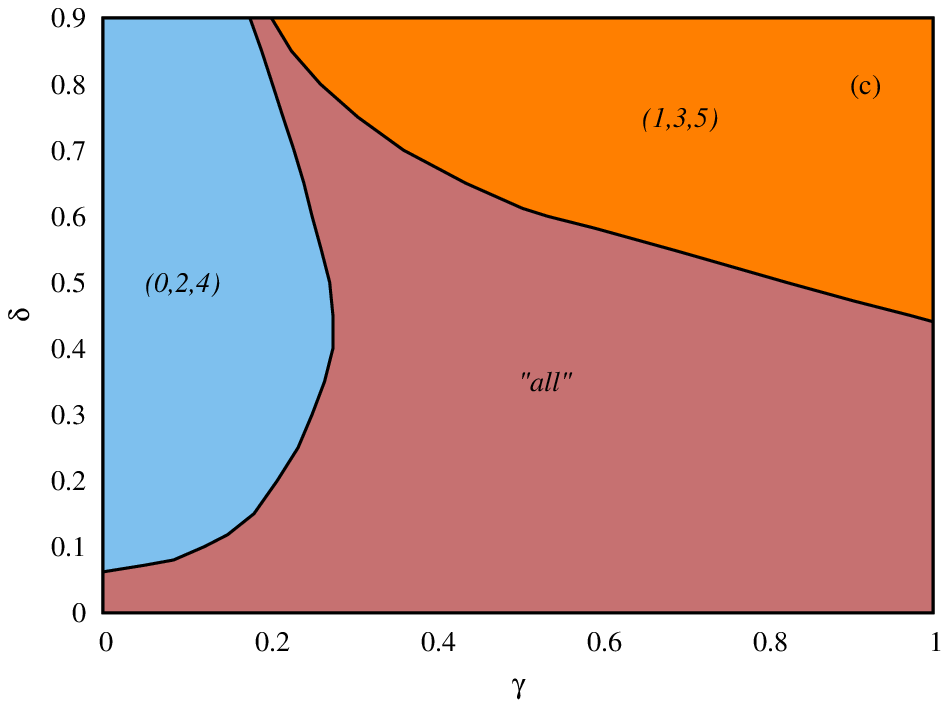,width=5.1cm}}
	\caption{\label{inter}Phase diagrams when interaction between rival triplets is intermediate ($\beta=0.5$). Horizontal axis shows the $\gamma$ blocking strength while vertical axis denotes the $\delta$ extra value of inner invasion in the even group. Panels~(a) to (c) respectively show cases when the inner invasion strength in the odd-labeled group is strong ($\alpha=0.7$), moderate ($\alpha=0.2$), or very weak ($\alpha=0.1$). Light blue (orange) marks the parameter area where even (odd) species prevail, while pink shows the area where all six species coexist. Dashed (solid) lines denote the position of discontinuous (continuous) phase transition points.}
\end{figure*}

In the following we survey the situation when the interaction between the competing triplets is intermediate. Again, as previously, we distinguish three cases where the inner invasion of odd-labeled triplet is fast, moderate, or very weak. The representative phase diagrams are shown in Fig.~\ref{inter}. When the inner rotation is fast, displayed in panel~(a), the system behavior is similar to those we observed for large $\beta$. Namely, ({\it 1+3+5}) species can form a solid winning alliance almost everywhere, the only exception is the low $\gamma$ - large $\delta$ corner where the faster rotation in the even-labeled triplet can compensate the shortage of blocking. When the invasion strength in the sound triplet is weaker, shown in panel~(b), we can observe a similar behavior previously found in Fig.~\ref{strong}(b). Namely, there is a parameter region where neither ({\it 0+2+4}) nor ({\it 1+3+5}) alliance is strong enough to beat the rival group hence all six species survive. There is, however, a qualitative difference compared to the previous case. In particular, the system behavior depends on the dynamical parameters on a more subtle way. If we fix $\gamma$, for instance, and only increase the value of $\delta$ then we can detect some non-monotonous behavior: starting from ({\it 1+3+5}) phase we can enter to ``{\it all}'', after ({\it 0+2+4}), followed by ``{\it all}'' again, and finally arriving back to ({\it 1+3+5}) phase again. 

This system behavior has serious consequence. Previous results obtained in other models and those we presented above for fast triplet rotations confirmed each other and suggest the general finding that faster inner invasion in a triplet is always beneficial to the alliance. This is partly true here because we can reach the ({\it 0+2+4}) phase by solely increasing $\delta$. But increasing this parameter further has a negative effect on the fitness of this loop. In short, a faster rotation becomes harmful to this alliance here. Before discussing this curious behavior more deeply, we first present the remaining low-rotation case, shown in panel Fig.~\ref{inter}(c). Since we decreased $\alpha$, the general rotation speed decayed in the ({\it 1+3+5}) group. As a consequence, the domain of this phase has shrunk in the phase diagram by giving space for ``{\it all}'' phase. But in contrast to Fig.~\ref{strong}(c), the orange phase does not disappear fully because we have smaller $\beta$, hence the large loop containing all six species is not as effective anymore as for large $\beta$ values. 

\begin{figure*}
	\centerline{\epsfig{file=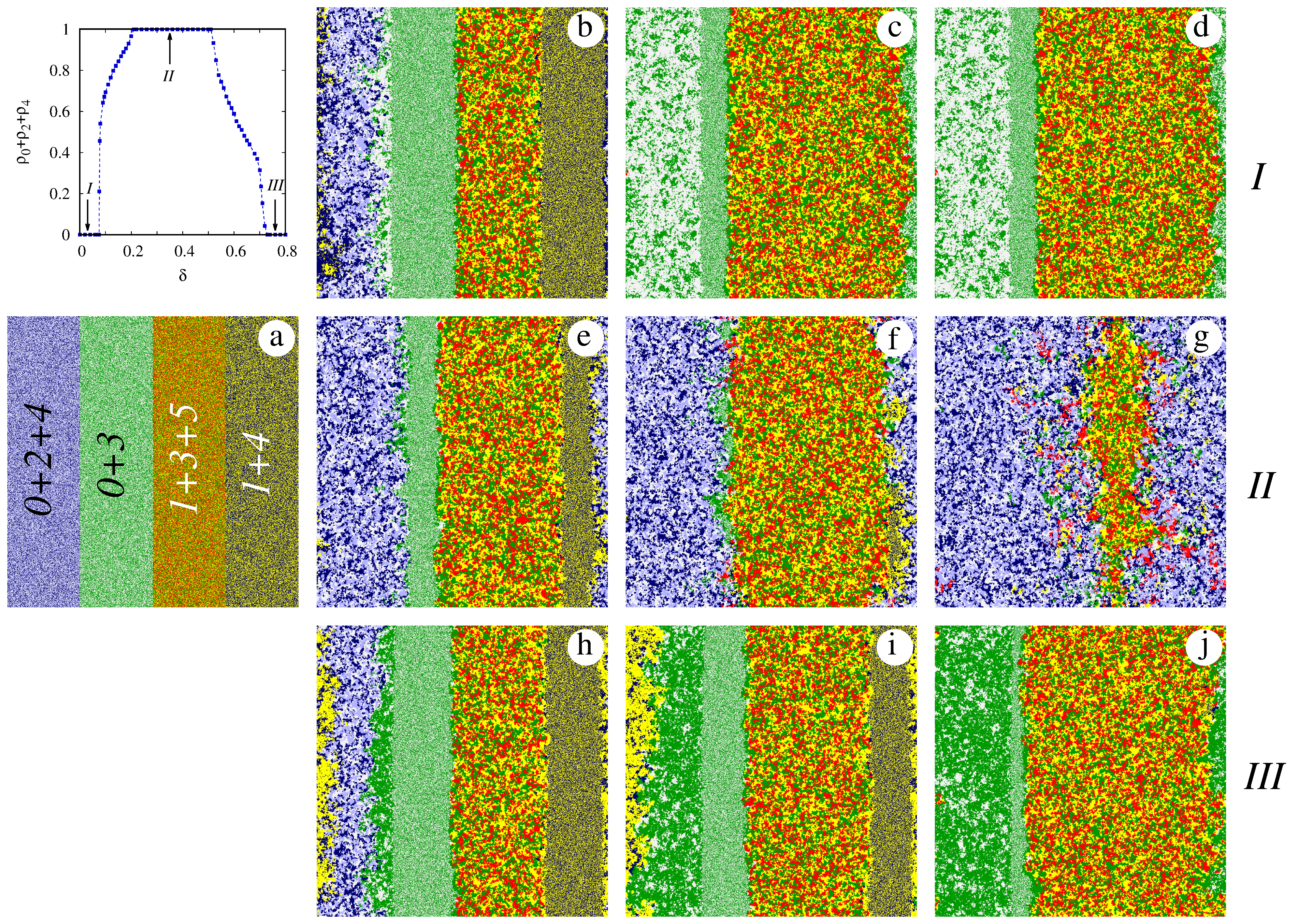,width=15.1cm}}
	\caption{\label{big}Reentrant phase transitions when rotation intensity is changed in the blocked triplet. Panel in the top-left corner shows the order parameter in dependence of $\delta$ at $\beta=0.5$, $\alpha=0.2$, $\gamma=0.25$. Arrows show the positions of the $\delta$ values where we launched the evolution from a specific initial state shown in panel~(a). The diverse evolution of patterns in Cases~$I$, $II$, and $III$ are shown in the rows. Final destinations, which are ({\it 1+3+5}), ({\it 0+2+4}), and ({\it 1+3+5}) phase again, are not shown. The color code of species is identical those used in Fig.~\ref{model}. The applied linear system size is $L=800$. Further details can be found in the main text.}
\end{figure*}

The top-left panel of Fig.~\ref{big} illustrates quantitatively the above mentioned curious re-entering phenomenon. Here we plotted the $\rho_0+\rho_2+\rho_4$ order parameter in dependence of $\delta$ at fixed $\beta=0.5$, $\alpha=0.2$, $\gamma=0.25$ parameter values. It is an important observation when we monitor the spatial battle of triplets that the front separating the competing domains is always fluctuating, hence patches containing only two neutral species can emerge easily. As a result, triplets do not simply fight each other, but they also compete with other solutions. The latter has particular role in understanding the unexpected phenomenon. To demonstrate it we present the evolution of spatial pattern in three different cases as marked by arrows in the mentioned panel. In all three cases the evolution starts from a prepared initial state where triplets are separated by domains containing neutral species. This common starting state is shown in Fig.~\ref{big}(a) where we indicate those species who are present in a specific stripe. As a technical note, for better visibility we only present two of the possible neutral pairs here, but similar pattern formation could be observed in the presence of ({\it 2+5}) pair solution. We first discuss Case~{\it I} obtained at $\delta=0.05$. In agreement with the standard expectation about defending alliances, the sound ({\it 1+3+5}) triplet beat both ({\it 0+3}) and ({\it 1+4}) duplets. Here both species {\it 0} or species {\it 4} can be considered as an intruder which is defeated by the rotating triplet. The same is not true for ({\it 0+2+4}) triplet because the blocking mechanism prevents them to defend effectively against ({\it 0+3}) or ({\it 1+4}) domains. After 1200 MC steps, as it is illustrated in panel~(b), yellow-blue and green-white patches emerge in the bulk of ({\it 0+2+4}) domain, signaling the negative consequence of blocking. After 5800 MC steps, shown in panel~(c), the latter formation disappears and the sound ({\it 1+3+5}) triplet gradually prevails. We stress that the latter formation are capable to dominate the alternative triplet directly, as it was illustrated in all previous case when $\delta$ is too small. As expected, the relation of triplets changes by increasing $\delta$ because a faster inner rotation makes blocked ({\it 0+2+4}) formation viable. It happens in Case~{\it II} shown in the middle row. As panel~(e) illustrates, taken after 5000 MC steps, both triplets can grow on the expense of domains formed by duplets. Panel~(f) depicts the event, after 8500 MC steps, when triplets meet and the superiority of even-labeled triplet over odd-labeled triplet becomes evident. The last row of Case~{\it III} shows when the inner rotation in the blocked triplet is almost maximal, which has a fatal consequence on the vitality of this even-labeled alliance. Panel~(h), obtained after 2000 MC steps, illustrates that duplet of neutral pairs becomes stronger again. Importantly, the strength of blocking mechanism is identical to all three cases, the only difference is the inner invasion strength in the blocked group. Let us consider, for instance, the battle of ({\it 0+2+4}) and ({\it 0+3}). While {\it 3}$\to${\it4} invasion remains intact, the {\it 2}$\to${\it 3} invasion is partly blocked. In this way the the former process has a larger probability if the inner rotation in the triplet is fast enough. It simply means that too fast rotation could be harmful for blocked triplet against two-member domains and they go extinct after 5000 MC steps, as shown in panel~(i). On the other hand, the sound triplet still dominates the two-member domains, hence odd-labeled group will be the final victor, no matter they would be weaker against the even-labeled group in a direct interaction. Summing up, the final destinations are identical in Case~{\it I} and in Case~{\it III}, but the temporarily emerging two-member patches have a critical role on defeating the even-labeled triplet in Case~{\it III}. While the odd-labeled trio is capable to beat the even-labeled trio directly in Case~{\it I}, this is not true in the alternative case. Here it needs the help of neutral pairs who can beat the blocked trio because of their too fast inner rotation. But later the neutral pair solutions give space for the rival triplet. This dynamics recalls ``the Moor has done his duty, the Moor may go'' effect previously observed in a different system modeling a social dilemma situation~\cite{szolnoki_pre11b}.

\begin{figure*}
	\centerline{\epsfig{file=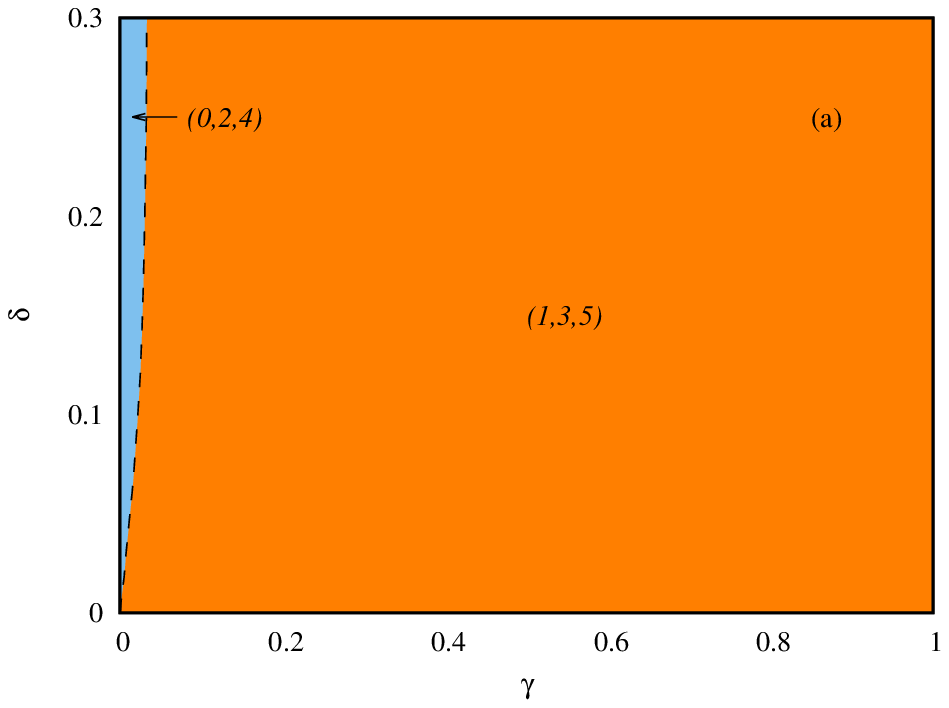,width=5.1cm}\epsfig{file=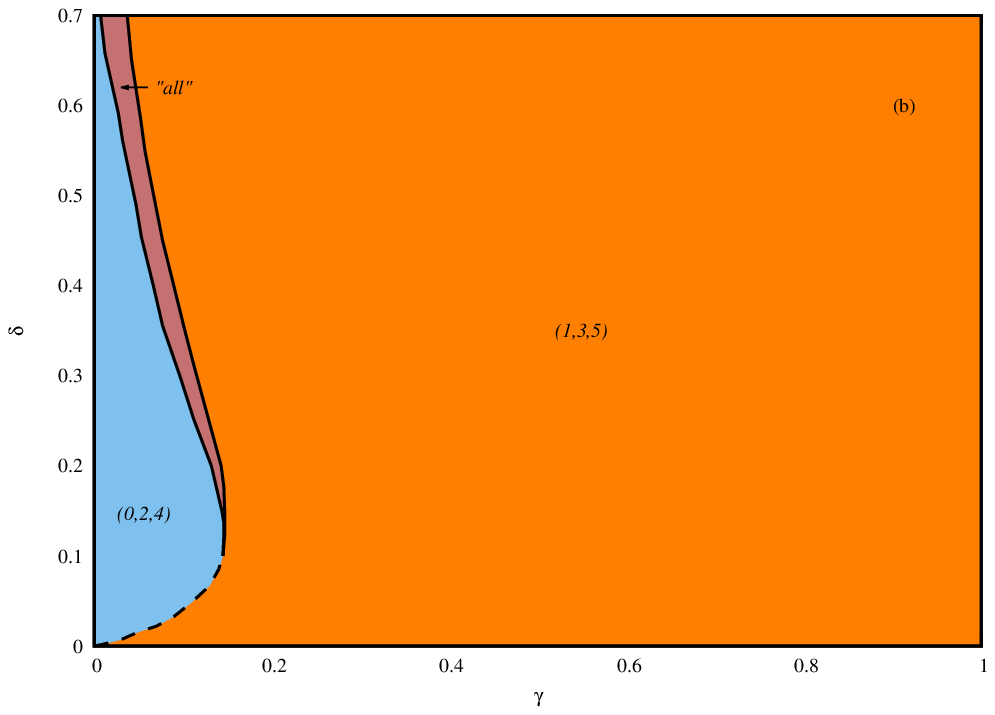,width=5.1cm}\epsfig{file=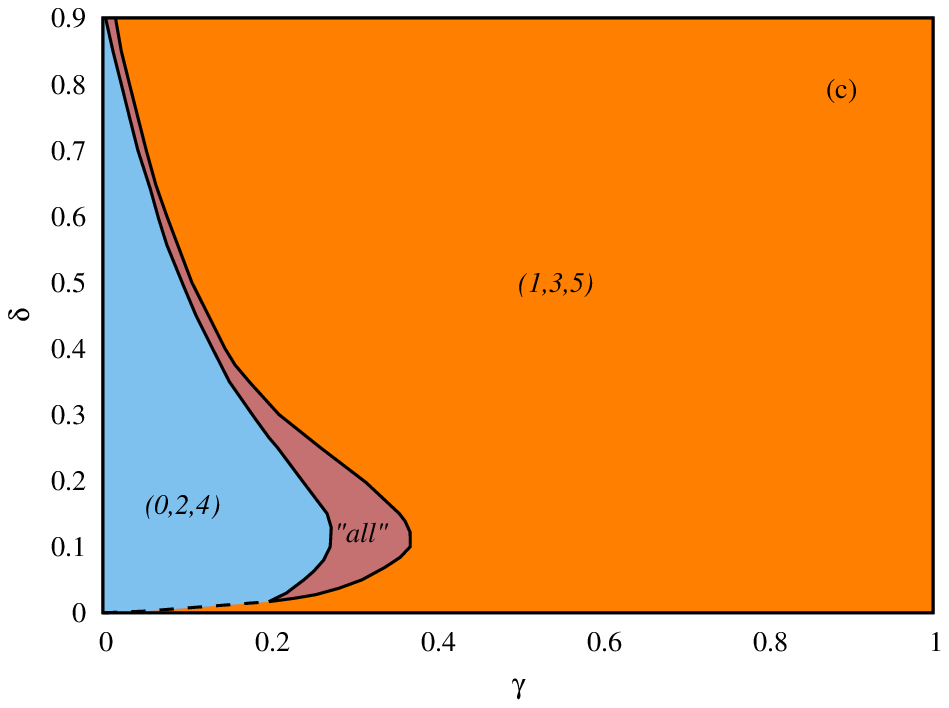,width=5.1cm}}
	\caption{\label{weak}Phase diagrams for weak interaction between the rival triplets ($\beta=0.2$). Horizontal axis shows the $\gamma$ blocking strength while vertical axis denotes the $\delta$ extra value of inner invasion in the even group. Panels~(a) to (c) respectively show cases when the inner invasion strength in the odd-labeled group is strong ($\alpha=0.7$), moderate ($\alpha=0.3$), or weak ($\alpha=0.1$). Light blue (orange) marks the parameter area where even (odd) species prevail, while pink shows the area where all six species coexist. Dashed (solid) lines denote the position of discontinuous (continuous) phase transition points.}
\end{figure*}

To complete our study we briefly summarize our results obtained for the case when interaction between the competing triplets is weak. As previously, panels of Fig.~\ref{weak} outline the typical behaviors for cases when the general rotation speed among triplet members is large, intermediate or slow. Based on our previous observations the system behavior can be interpreted in the following way. Similarly to those we detected for other $\beta$ values, when the general inner rotation is high, shown in panel~(a), the competition between the triplets is always clear: either even- or odd-labeled group win. The only difference is the consequence of blocking is more serious now. Consequently, the drawback of ({\it 0+2+4}) group can be compensated just for very small $\gamma$ values. The re-entrance phenomenon can be observed both in panel~(b) and panel~(c), respectively representing intermediate and slow inner rotation. The area of ``{\it all}'' phase, however, is significantly smaller than for other $\beta$ cases discussed previously. The latter feature can be explained by the small $\beta$ value which generally makes the large six-member loop fragile. We must note, however, that the $\beta \to 0$ limit provides a qualitatively different situation where all six species coexist again. In the latter case there is no interaction between the competing triplets anymore, hence they can form two there-member stable solutions independently. But this phase is different from the one we call as ``$all$'' here because the latter is based on a cyclic dominance among all six competitors.

\section{Conclusion}
\label{conclusion}

Different aspects of cyclically dominant dynamical systems have been studied intensively in the last years~\cite{menezes_csf23,park_c22,szolnoki_csf20,serrao_epjb21,avelino_epl18,roman_jtb16}. By choosing a similar system, the key question in our present work is to clarify how a blocking mechanism affects the vitality of the involved formation. Importantly, such blocking, when the interaction between two actors is influenced by the vicinity of a third partner, is a common phenomenon in microbiological systems or even in human societies~\cite{alvarez-rodriguez_nhb21}.

While this kind of blocking may stabilize a coexistence in a three-member system~\cite{bergstrom_n15}, there are two reasons why such blocking could be harmful for a cyclic loop in a more complex situation. The first one is plausible because members may impede each other to invade successfully an external party, which can be detrimental for the whole alliance. The other one is the consequence of retarded inner rotation. In this case the average size of domains formed by the members of the loop grows. And the increase of local homogeneity is always disadvantageous for a defending alliance which is based on cyclic dominance.

By using a minimal model, where two three-member loops compete, we demonstrated that the above described drawbacks can be compensated by a faster inner rotation. In this case the blocked triplet can dominate the sound formation where blocking is not applied. Naturally, if the blocking probability is too large then the untouched formation will be the victor and the transition between these solutions is discontinuous. This observation remains valid independently of the interaction strength between the triplets. If the general invasion withing the alliance is weak then a higher loop, involving all six species, can win resulting in the stable coexistence of all competitors.
 
When the interaction between the triplets is moderate or weak then we could detect interesting reentrant phase transitions by solely changing the inner invasion speed in the blocked triplet. First, if rotation becomes intensive then blocked triplet become dominant. If we increase the rotation speed further, however, the original block-free triplet wins again. The explanation of this unexpected behavior is based on an special ``Moor-effect'' where a third type of solution brings the final victory to the block-free triplet. In a ``Moor-effect'', when originally two solutions compete, a third solution can beat one of them and later it gives space to the another competitor~\cite{szolnoki_pre11b}. In our present case when two triplets compete, their frontier serves as a birthplace of alternative solutions. Namely, there is always a chance that a two-member solution of neutral species emerges. If the rotation speed is too high in the blocked triplet then such two-member solution, formed by a triplet member and an external species, can be more effective: while the invasion of external party is not blocked, it can meet more easily with its internal predator who cannot beat it due to blocking. As a consequence, the two-member formations invade the blocked triplet, leaving only the alternative triplet alive. Importantly, the defensive mechanism still working in the sound triplets, hence they eventually prevail. No matter they would be weaker in a direct comparison against the blocked triplet.

As a general conclusion, our system provides a nice example that the consequences of dynamical rules determining the vitality of a solution are more subtle than we originally thought. Therefore further research is necessary to establish solid principles to understand such ecological systems more deeply.
  
\begin{acknowledgments}
A.S. was supported by the National Research, Development and Innovation Office (NKFIH) under Grant No. K142948. X.C. was supported by the National Natural Science Foundation of China (Grant Nos. 62036002 and 61976048).
\end{acknowledgments}

\end{document}